# Prototype Design of Singles Processing Unit for the Small Animal PET

**P. Deng,**[a,b] **L. Zhao,**[a,b,*] **J. Lu,** [a,b] **B. Li,** [a,b] **R. Dong,** [a,b] **S. Liu**[a,b] **and Q. An** [a,b]

[a] *State Key Laboratory of Particle Detection and Electronics,*
*University of Science and Technology of China, Hefei 230026, China*
[b] *Department of Modern Physics, University of Science and Technology of China,*
*Hefei 230026, China*
*E-mail*: `zlei@ustc.edu.cn`

ABSTRACT: Position Emission Tomography (PET) is an advanced clinical diagnostic imaging technique for nuclear medicine. Small animal PET is increasingly uesd for studying the animal model of disease, new drugs and new therapies. A prototype of Singles Processing Unit (SPU) for a small animal PET system was designed to obtain the time, energy, and position information. The energy and position is actually calculated through high precison charge measurement, which is based on amplification, shaping, A/D conversion and area calculation in digital signal processing domian. Analysis and simulations were also conducted to optimize the key parameters in system design. Initial tests indicate that the charge and time precision is better than 3‰ FWHM and 350 ps FWHM respectively, while the position resolution is better than 3.5‰ FWHM. Commination tests of the SPU prototype with the PET detector indicate that the system time precision is better than 2.5 ns, while the flood map and energy spectra concored well with the expected.

[*] Corresponding author.

## Contents

## 1. Introduction

Position Emission Tomography (PET) is an advanced clinical diagnostic imaging technique for nuclear medicine. The principle of PET is to inject radiolabeled molecular contrast agents into an organism. Then agents are redistributed in the organism through various physiological processes. The PET agent emits a positron when it decays. The positronium annihilation caused by the positron produces two 511 keV gamma rays that travel in opposite directions. These ray signals are measured and coincided by high-resolution PET instrument. And the measured data are reconstructed to three-dimensional PET images by computers [1, 2].

The small animal PET is increasingly used in a variety of biomedical research, such as the animal model of disease, new drugs and new therapies [3-6]. For small animal PET, high spatial resolution and sensitivity are essential. To meet the high performance requirements, small animal PET is composed of high spatial resolution and sensitivity detector, high-speed high-precision electronics and proper image reconstruction algorithm.

This paper presents the design of a Singles Processing Unit (SPU) prototype, and it is aimed to be used in a high performance PET detector, which consists of lutetium-yttrium oxyorthosilicate (LYSO) crystal [7], Silicon Photomultipliers (SiPM), resistor network, and temperature sensors, shown in figure 1. The LYSO that have better time resolution and spatial resolution, is the most commonly used crystal for PET detectors at present [8, 9]. Because of its large gain, low working voltage (< 100 V), magnetic field compatible, good time resolution, SiPM is used as the light detector for this PET instrument [10, 11].

This PET instrument is actually a ring consisting of 12 detector modules. One detector module is made up of four detector blocks (B1, B2, B3, and B4 in figure 1). Each detector block is composed of 23×23 LYSO crystals, 2 layers of SiPMs, and 2 resistor networks.

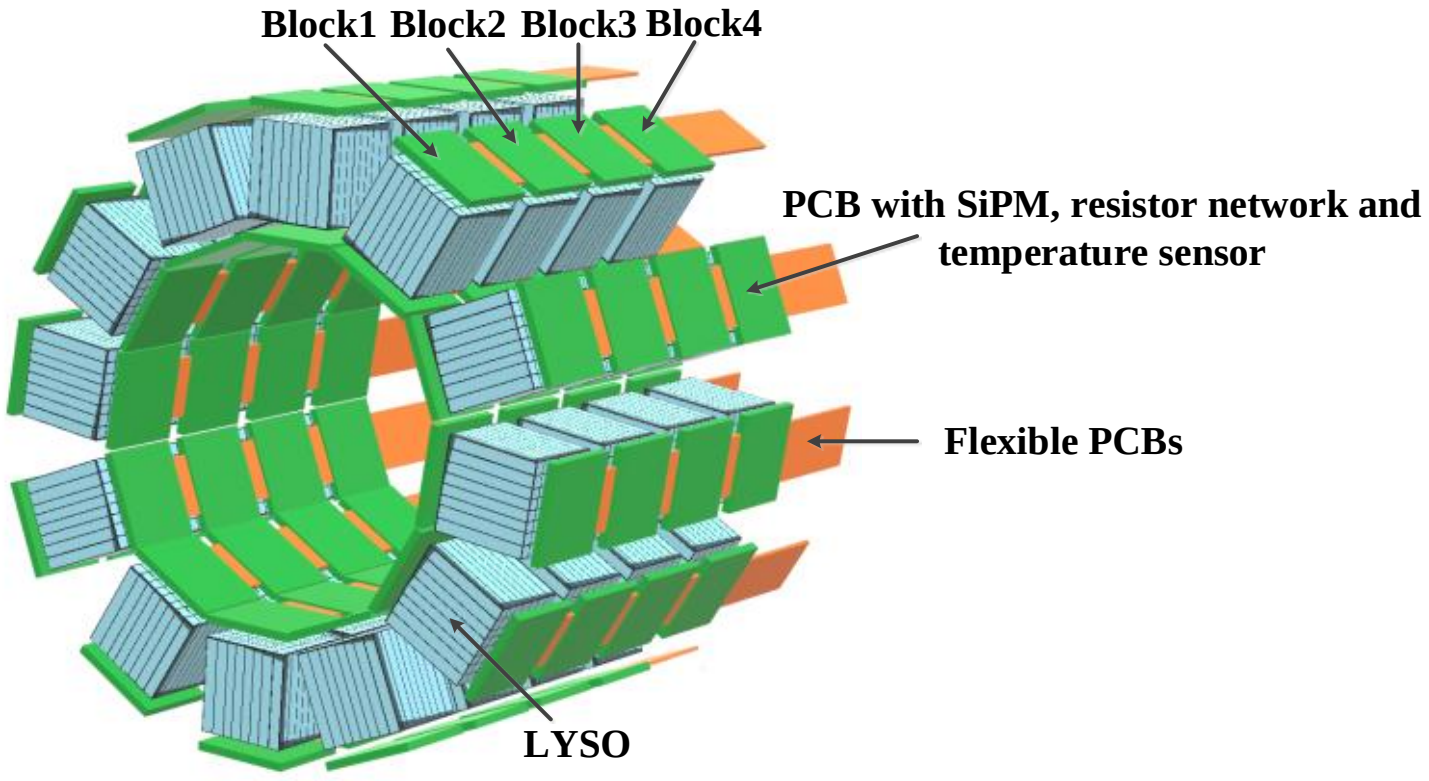

Figure 1. A high small animal PET detector.

To process the PET signal and get the position and energy information, high performance readout electronics is indispensable. The structure of the whole electronics is shown in figure 2, and it consists of Front Amplifier Boards (FABs), SPUs, Coincidence Processing Unit (CPU), SiPM Supply Boards (SSBs), Clock and Synchronize Board (CSB), Power Supplier (PS), 10G Ethernet Switch and computer. The PET detector senses the gamma rays and generate electrical signals. Then the signals are amplified by FAB and then fed to SPU to obtain the required position, energy, and time information, and the result data are finally transferred to CPU module by a 10G Ethernet Switch. In the CPU module, coincidence among the data is conducted and the reorganized data are finally sent to computer to reconstruct the PET images.

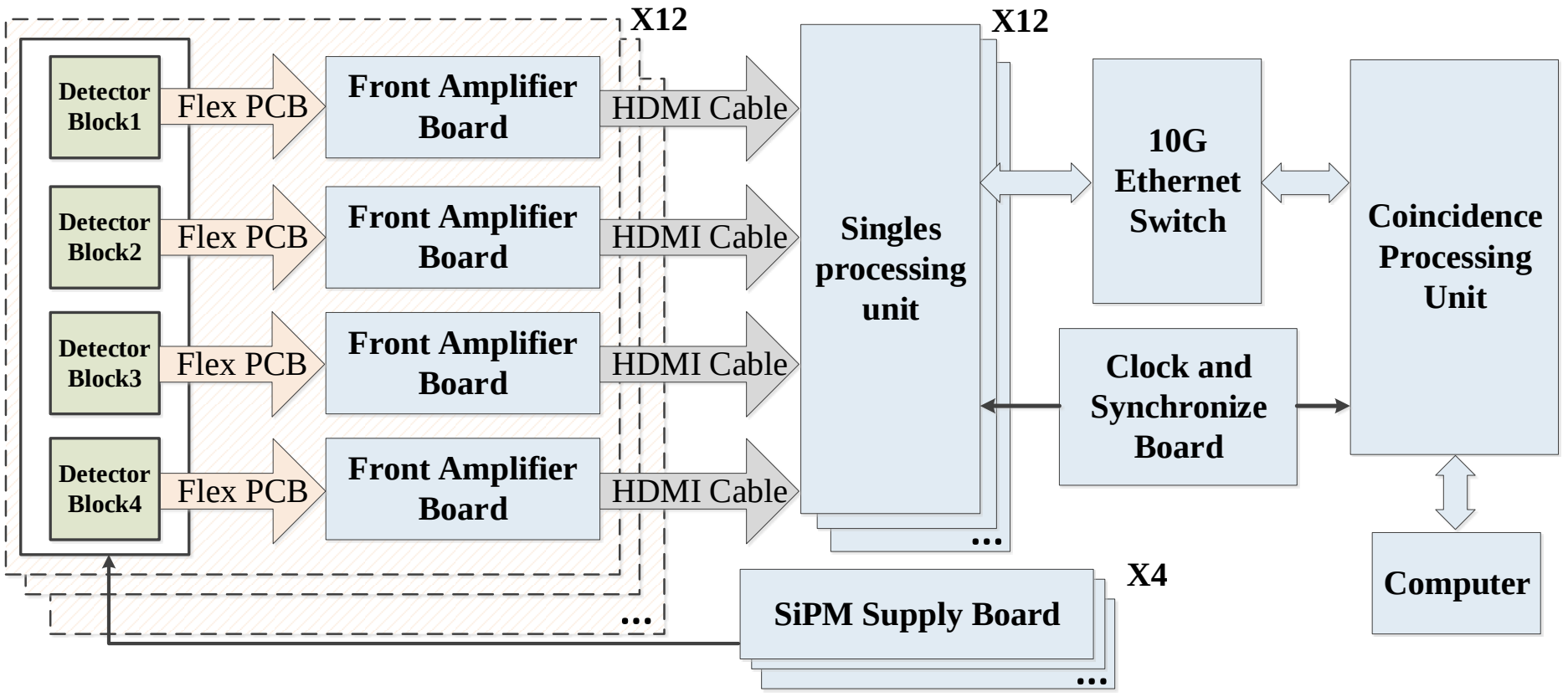

Figure 2. The whole readout electronics of the small animal PET.

In this paper, we focus on design of the SPU module prototype for this small animal PET instrument. There are 32 channels in each SPU, corresponding to 4 detector blocks, and high-resolution time and charge measurement is required. Table 1 shows the performance specifications of the SPU.

| Performance specification | Requirement |
|---|---|
| Charge precision | ⩽ 0.5 % FWHM (Full Width at Half Maximum) |
| Time precision | ⩽ 500 ps FWHM |
| Position precision | ⩽ 0.5 % FWHM |

Table 1. The performance requirements of SPU.

## 2. SPU Design

### 2.1 Architecture of the SPU module

The architecture of the SPU module is shown in figure 3. As one of the key parts in the electronics, the SPU receives the signals from the FAB through 8 HDMI connectors. The input signal is differential, the leading time is about 50 ns, and the summation of 8 signals from one detector block is about 2 V. In the SPU, the input signal is amplified and then split into two paths: one passes through shaping circuits, followed by an analogue to digital conversion(ADC), and the other is fed to a summation circuits (8 channels to 1). The data from the ADCs are used to obtain the charge information based on area calculation, while the output from the summation circuits is further fed to discriminators followed by FPGA-based Time-to-Digital Converters for time measurement. The corresponding digital signal processing logic is integrated in one Field Programmable Gate Array (FPGA). According to the application requirement, three working modes are implemented in the FPGA, including regular processing mode, flood map mode and energy spectrum mode logic. Finally, the data results are sent to the CPU module by the Ethernet interface based on the UDP protocol.

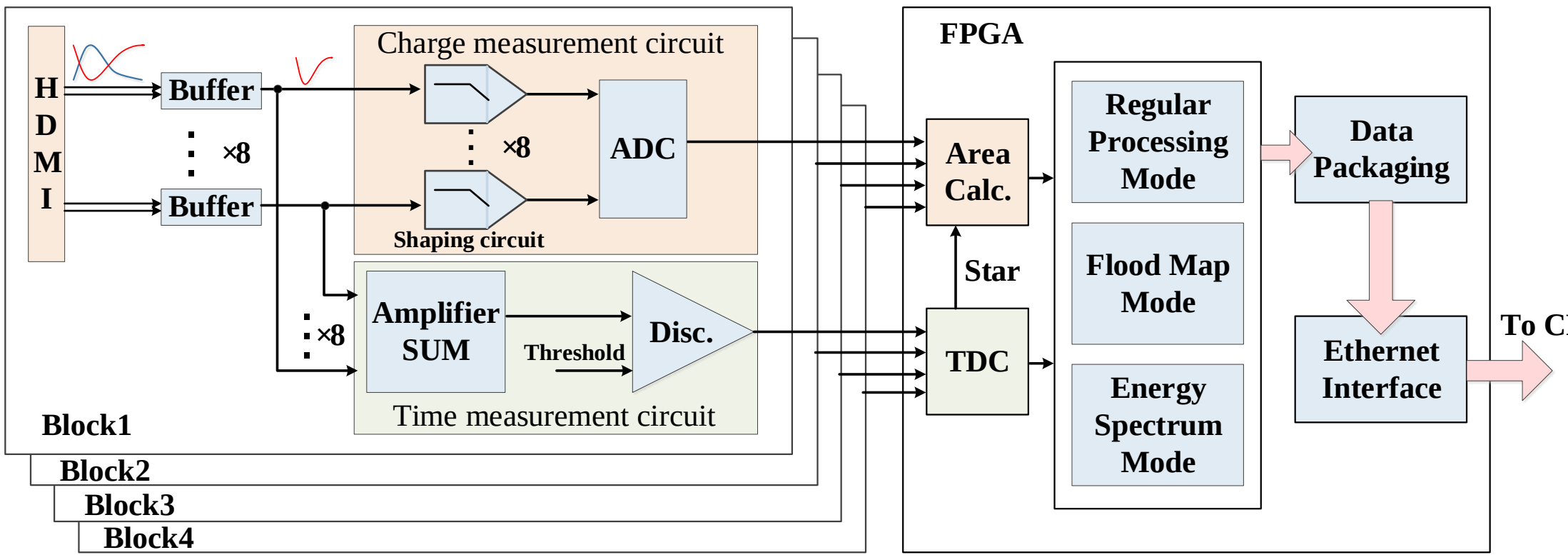

Figure 3. The architecture of the SPU module.

### 2.2 Design of the charge measurement circuit

As shown in figure 3, the charge measurement circuit consists of the analogue signal processing and the digital signal processing logic in FPGA.

#### 2.2.1 Amplification, shaping circuit, and ADC

Figure 4 shows the analogue signal processing of one channel for charge measurement. For each input signal from the FAB, the 50 Ω resistor R0 is used to achieve impedance matching, and R and the TVS (Transient Voltage Suppressors) diode are used for over charge protection. Then each pair of differential input signals is amplified by a high-speed amplifier AD8099 from ADI Corporation [12] (A1 in figure 4). The output of A1 is split into two paths: one is fed to the following charge measurement circuits, while the other is fed to the time measurement circuits which will be presented in Subsection 2.3. The signal from A1 is shaped by an $RC^2$ filter (R1, C1, R2, C2 and A2 in figure 4). A2 (AD8057 from ADI Corporation [13]) is used to separate the two stages of $RC^2$. The differential amplifier A3 (AD8137 from ADI Corporation [14]) is used to convert the signal to a differential one, and match the signal amplitude to the FSR (full scale range) of the ADC AD9222 [15]. The sampling clocks for the ADC chips are from one PLL chip, which synchronize the sampling clocks with the input system clock from CSB module and for jitter cleaning.

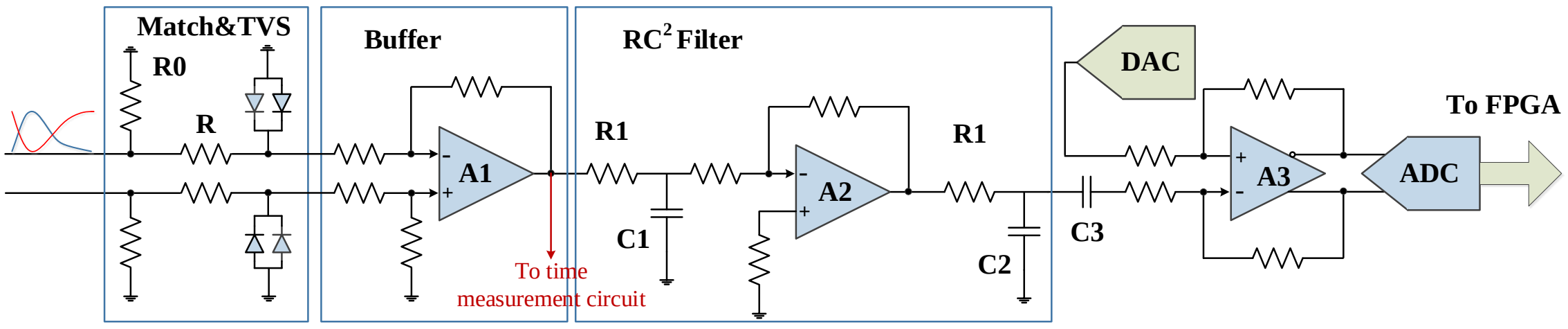

Figure 4. Analogue signal processing for the charge measurement.

### 2.2.2 Digital signals processing logic

The outputs of ADCs are fed to an FPGA device for charge calculation. In this design, we calculate the area of the digitized signal waveform to obtain the charge information. As shown in figure 5, the logic including ADC data deserialization logic and area calculation logic.

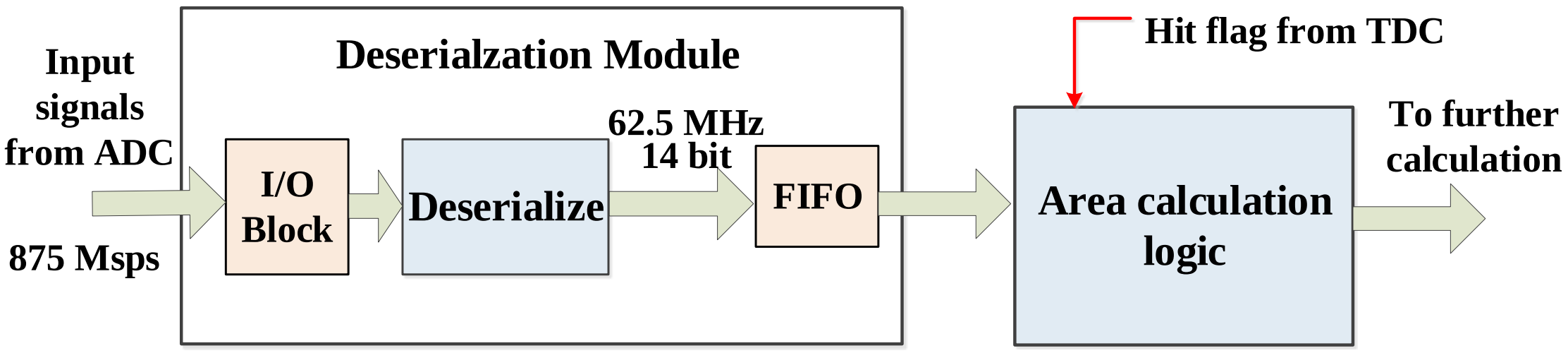

Figure 5. The structure of digital signals processing logic.

Since the ADC output data are 875 Msps serialized data streams, we need to deserialize the data to 62.5 Msps data with a data width of 14 bits. The structure of the deserialization logic is shown in figure 6. The data clock output (D_CLK), frame clock output (F_CLK) and data (DATA_IN) from ADC are fed to this logic block. To implement the 1:14 deserialization of ADC data, two groups (each consisting of two cascaded ISERDESE2 IP cores) are employed, marked as Group 1 and 2 in figure 6. In Group 2, DATA_IN are registered by the ISERDESE2 IP Cores with the clock D_CLK (62.5 MHz $\times$7= 437.5 MHz) working in Double Data Rate (DDR) mode, and then converted to 14 bits parallel data streams. To determine start bit in the original serial data stream, we use D_CLK to sample the other output clock of ADC F_CLK (62.5 MHz). By controlling the BITSLIP port of the ISERDESE2 Core, we can observe the changing of the output data. When the output of Group 1 is "11111110000000", it means that we get the correct output data from Group 2, since these two groups are tuned simultaneously. The output data are stored in a FIFO and then read out to the following area calculation logic block.

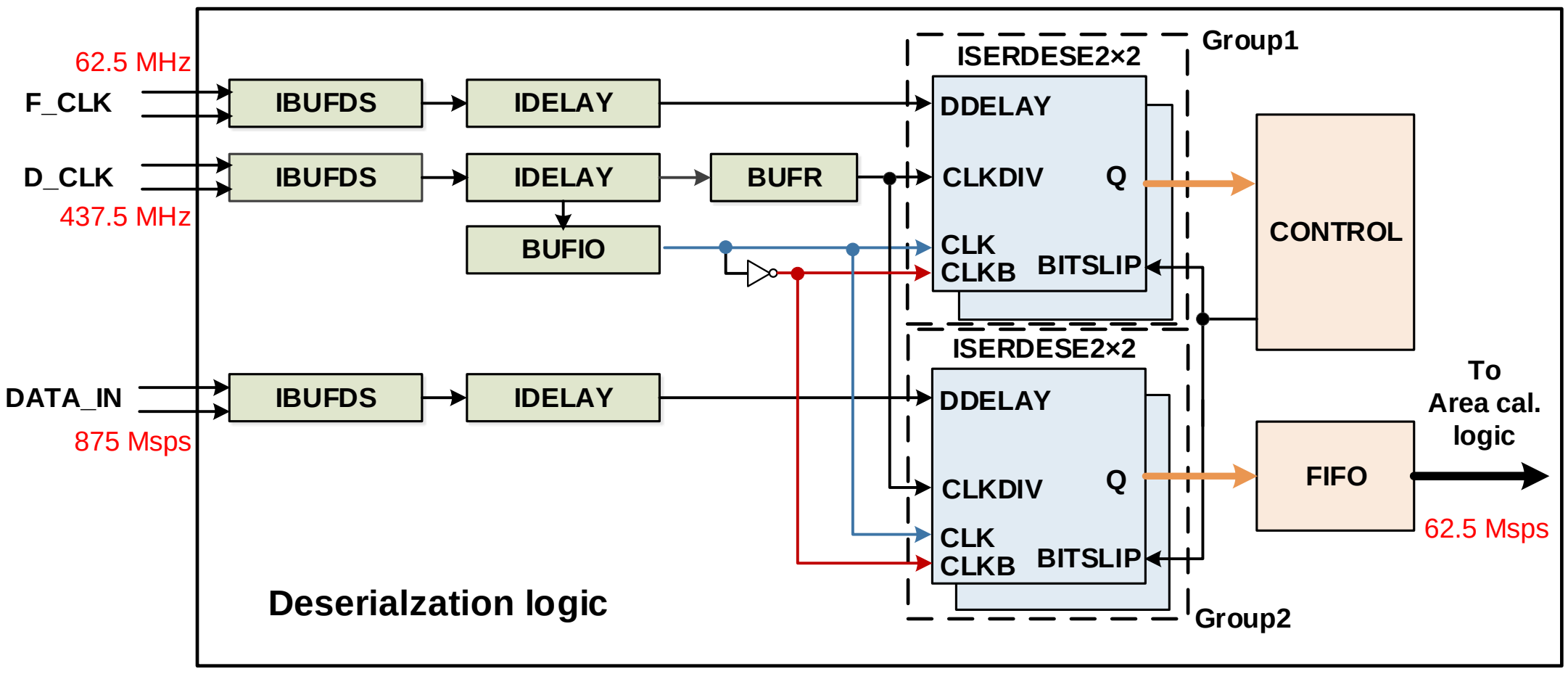

Figure 6. The structure of the ADC data deserialization logic.

As for the digitized data, there are two possible methods to obtain the charge information. One is to directly detect the peak value of the digital signal waveform, and the other is to sum the amplitudes of the sample points (as shwon in figure 7, area S=sum($P_1$,...$P_N$)), i.e. area calculation. In the former, there exist peak detection error especially when the sampling speed is not high enough compared with the input signal frequency. This error can be decreased effectively by the latter, meanwhile the RMS noise would also be reduced through the summation processs (actually an averaging process).

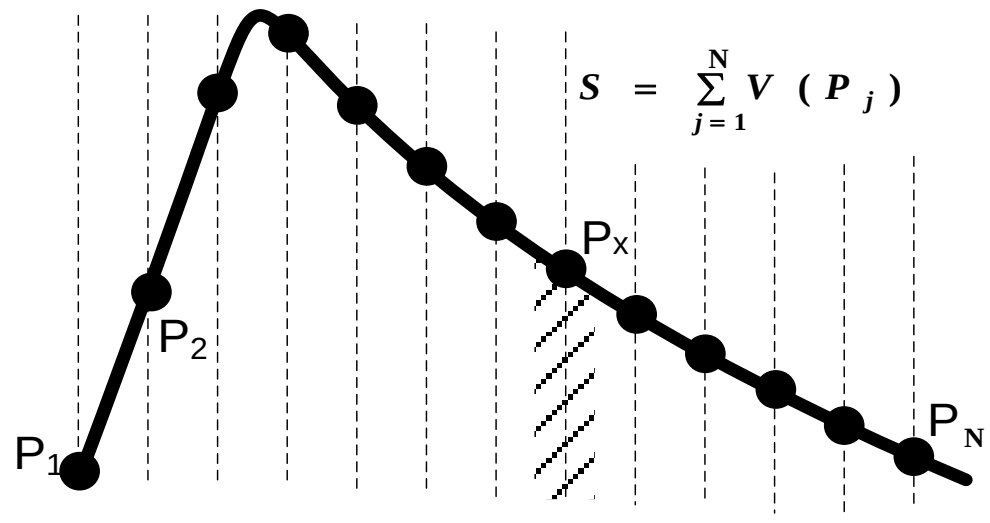

Figure 7. Diagram of the area calculation.

There are two factors that may influence the final charge measurement resolution: one is the number of samples (N) in the waveform added in the area S, and the other is the sampling frequency of the ADCs (marked as $f_S$). To evaluate their influence, we conducted simulations in the MATLAB platform. In the first simulation, we assume the sampling frequency is fixed to 100 MHz, and then we change the N from 1 to 100, and calculate the charge RMS resolution. This RMS uncertainty is caused by the uncertainty of the sampling time points on the waveform, and the maximum uncertain is actually $T_S$ (one sampling clock period, i.e. $1/f_S$=10 ns), as shown in figure 7 (the gray region around $P_X$). The simulation results in figure 8 indicate when N$\geqslant$55, the charge precision is better than 0.2 % FWHM, and we can observe that once N exceeds a certain value, the charge resolution is good enough, and cannot easily be enhanced.

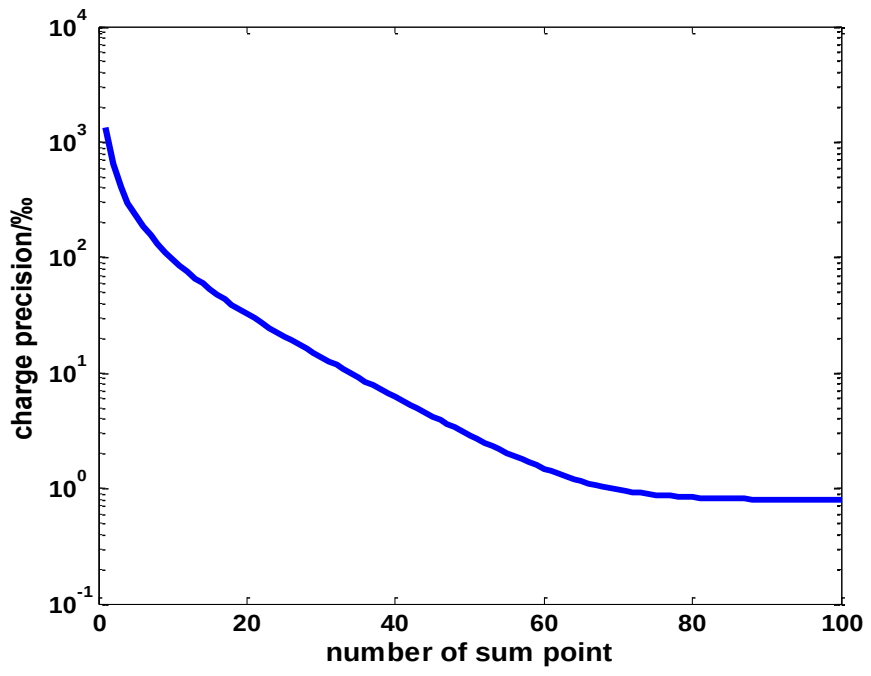

Figure 8. The charge precision when changing N.

As mentioned above, the other factor is the sampling speed. To estimate the overall effect of these two factors, we changed the values of them and obtained the simulation results presented in figure 9. As we can see in figure 9, when $f_S$ is around 60 MHz, and N⩾41, the charge resolution is better than 0.2 % FWHM, which is enough for our application. Of course, with a higher sampling speed and bigger N value, the resolution would be enhanced a little with higher power consumption, higher hardware cost and larger calculation dead time. According to the above analysis and simulation, optimization of the key parameters can be selected (N=62, $f_S$=62.5 MHz).

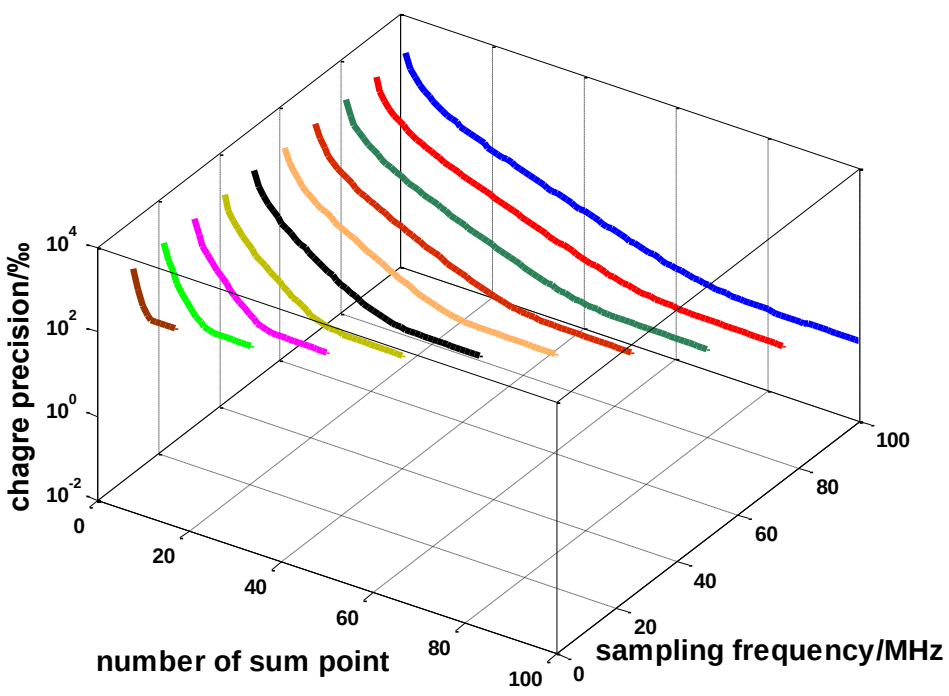

Figure 9. The simulation results when changing the two factors.

Figure 10 is the block diagram of the area calculation logic implemented in the FPGA. In this design, the hit flag for the calculation logic is generated in the TDC logic which will be presented in Subsection 2.3. To reduce the baseline fluctuation of the shaped signal after A/D conversion, we calculate the average value of the samples over several sampling clock periods before the actual start time of the signal, as shown in figure 10. Under the control of this logic, when the baseline calculation is finished, the digital data are switched into a summation logic, and before summation the baseline value is subtracted from each sample to get the correct value. The sample numbers that are included in the summation process (i.e. N in figure 10) are stored in a register, which can be modified by access from a remote PC.

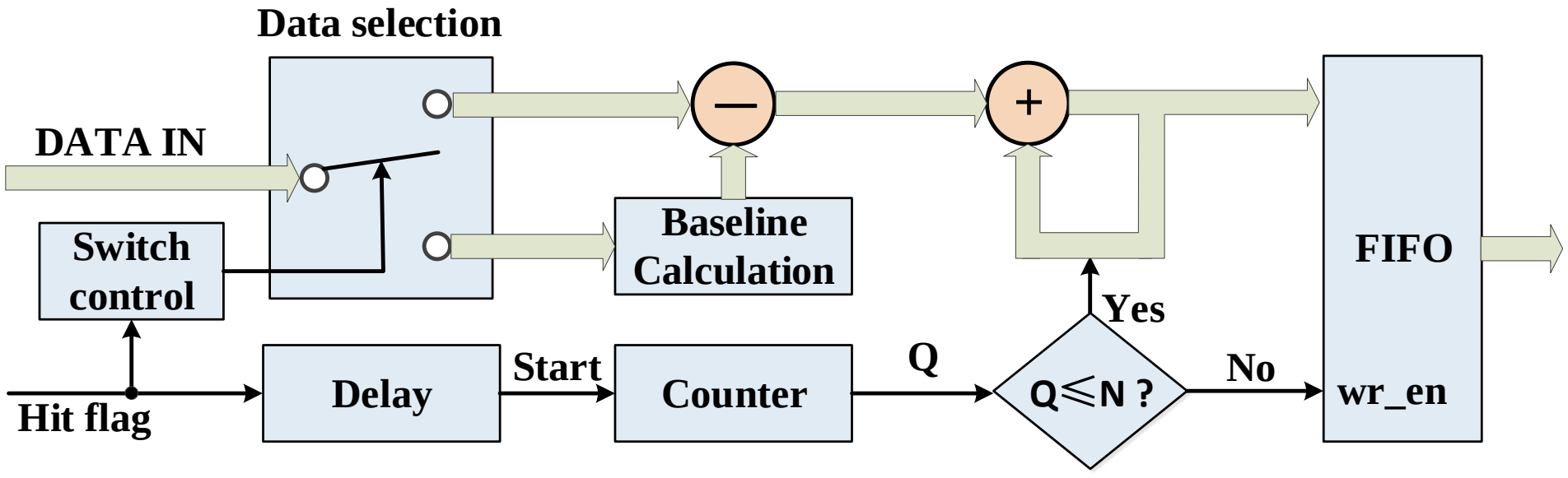

Figure 10. The block diagram of area calculation logic.

### 2.3 Design of the time measurement circuits

As shown in figure 3, the time measurement circuit consist of the analog circuits and an TDC logic in FPGA.

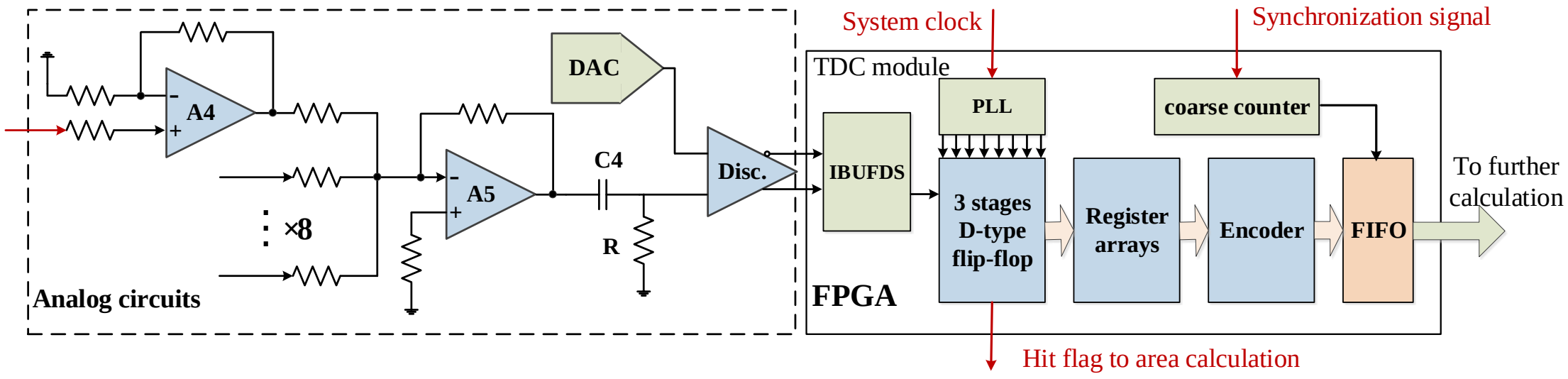

Figure 11. Time measurement circuits.

As shown in figure 11, the analog circuits receive the signal from A1 in figure 4, and signals of 8 channels are summed together to form one signal (with a constant peak value of around 2 V), which is fed to discriminator LMH7322 [16]. LMH7322 outputs LVDS signals to the TDC in the FPGA for final time digitization. Considering the signal baseline before discrimination, an AC coupling circuit (C4 and R in figure 11) is employed.

According to the time resolution requirement of 500 ps FWHM (RMS ~ 212 ps), the TDC is designed based on the multiphase clock interpolation method. The frequency of the main clock in the FPGA is multiplied through the internal PLL inside the FPGA by a factor of 6, and then 8 clock signals are generated for the TDC with a phase interval of 45 ° among them. Finally an equivalent bin size of around 330 ps can be obtained, and this corresponds to a maximum RMS error of 0.5 ×330 ps for time interval measurement between two TDC channels [17]. Considering the time results of the two TDC channels are not interrelated, the maximum RMS error of single channel can be obtained by dividing the 0.5 ×330 ps by $2^{1/2}$(i.e. 117 ps), the design can meet the requirement. As shown in figure 11, in order to address the metastability issues, three stages of D-type flip-flops are employed to latch the discriminator output. The register arrays store the fine time results of TDC results temporarily. When the system clock arrives, the fine time value (48 bits) is sent to the encoder. The encoder translates this 48 bit thermometer code and generates a 6 bit fine time. The counter (with a width of 37 bits) in figure 11 is used for coarse time measurement, and the time measurement range of above 30 min. Finally, the fine time and the coarse count value are combined together and stored in a FIFO.

## 3. Test result

### 3.1 Laboratory tests of SPU

To evaluate the performance of this prototype SPU, we conducted a series of tests in the laboratory. We used an arbitrary signal source (AFG3252 from Tektronix Inc.) to generate the input signal according to the waveform of PET detector output signal captured by high speed oscilloscope.

Figure 10 shows a typical waveform sampled by the ADC with an amplitude of 2 V.

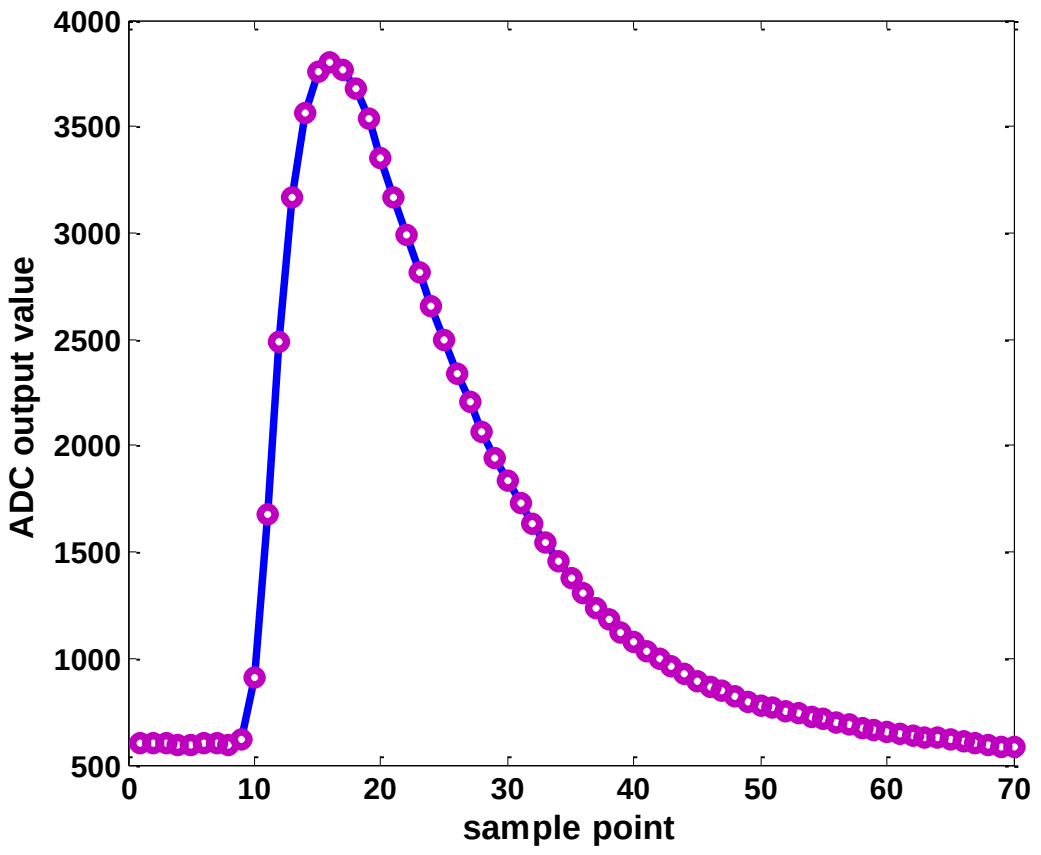

Figure 10. A typical waveform sampled by the ADC.

Figure 11 (a) shows a typical histgoram of the charge measurement results with the signal amplitude of 2 V; Figure 11 (b) are the charge resolution test results for all the 32 channels. The resutls indicate that the charge resolution of 32 channels are all about 2.9‰ Full Width Half Maximum (FWHM), which is good enough for the application.

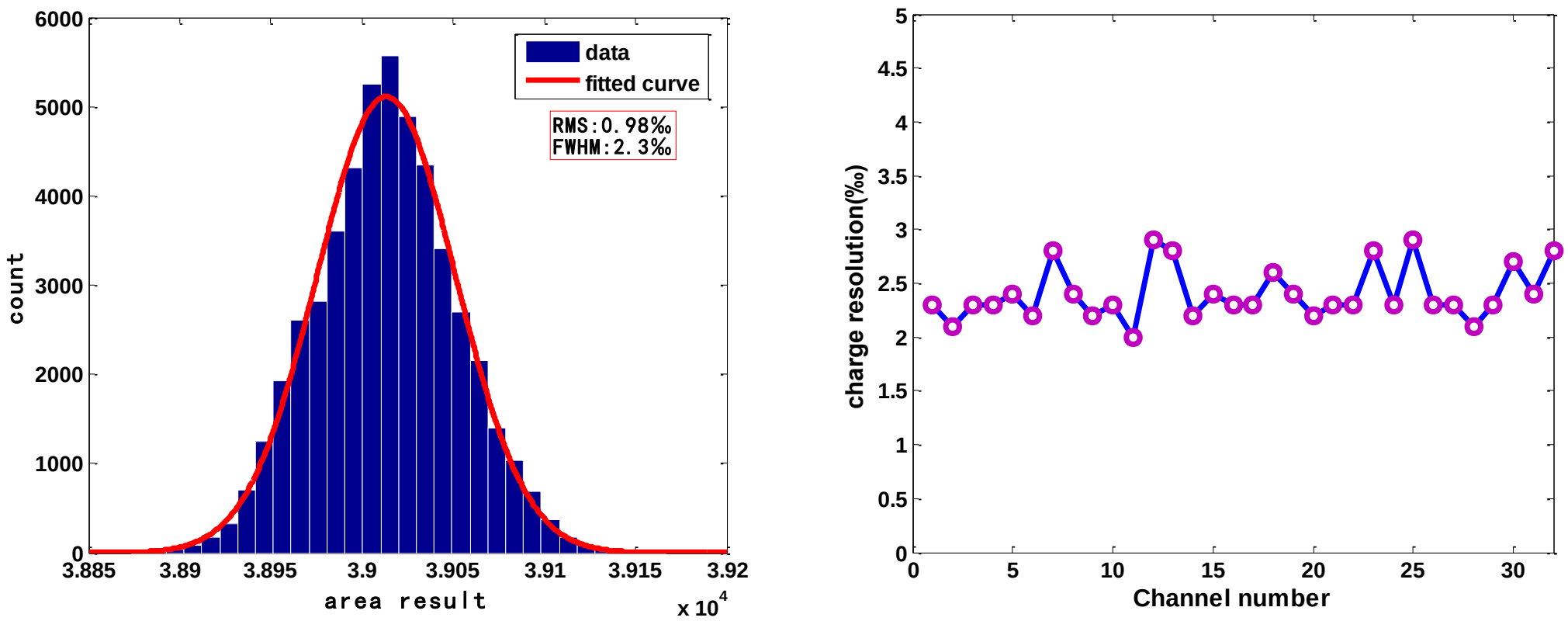

(a) (b)

Figure 11. Charge measurement results, (a) a typical histgoram, (b) all the 32 channels.

In the time measurement performance test, we evaluate the resolution based on the “cable delay test” method [18]. Considering that each 8 channels are summed together before being fed to discriminators, there are a total of 4 time measurement channels in one SPU. We directly calculate the difference of the time measurement results between two channels (in the test, Channel 1&2 are in one pair, and Channel 3&4 are in the other pair), and can finally get the performance of one single channel. Figure 12 (a) and (b) are the histograms of the time difference measurement result divided by $2^{1/2}$ for Channel 1 &2, Channel 3&4. We also changed the time difference between the two channels in one pair, and evaluate the time performance. The results in Figure 13 indicate that the time resolution is around 301 ps FWHM for all channels.

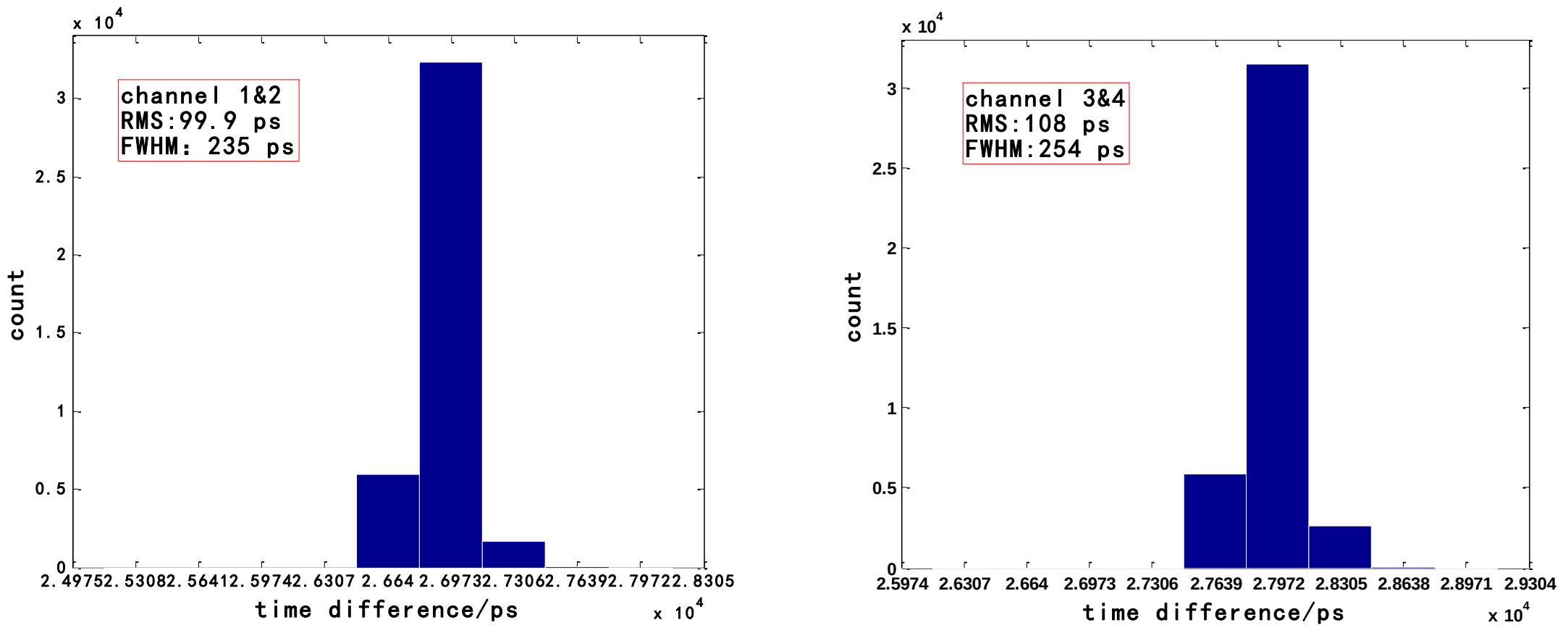

(a) (b)

Figure 12. Histograms of time measurement results, (a) Channel 1&2, (b) Channel 1&2.

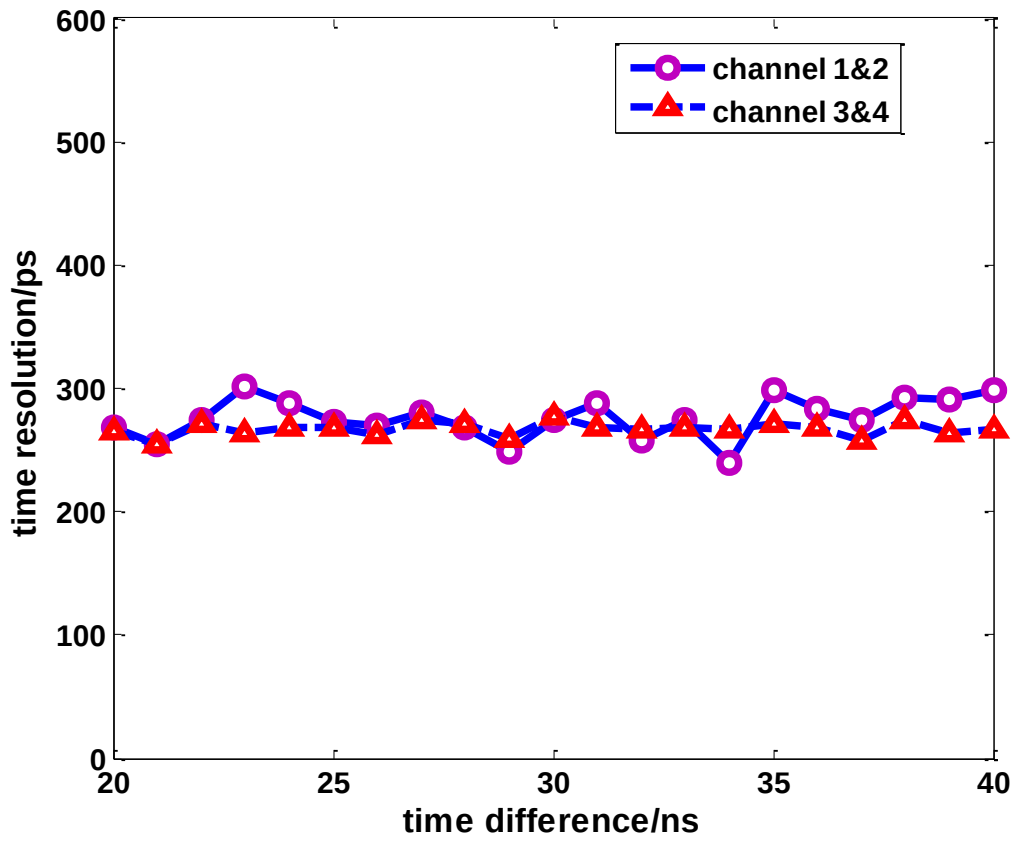

Figure 13. Time resolution with different time defference for all channels.

As for position measurement, the position information is actually calculated from the amplitude of the 8 channels in one detector block, as in (1) and (2).

$$x = \frac{1}{2}\left(\frac{A_1+D_1}{A_1+B_1+C_1+D_1}\right) + \frac{1}{2}\left(\frac{A_2+D_2}{A_2+B_2+C_2+D_2}\right) \tag{1}$$

$$y = \frac{1}{2}\left(\frac{A_1+B_1}{A_1+B_1+C_1+D_1}\right) + \frac{1}{2}\left(\frac{C_2+D_2}{A_2+B_2+C_2+D_2}\right) \tag{2}$$

where $A_1$, $B_1$, $C_1$, $D_1$, $A_2$, $B_2$, $C_2$, and $D_2$ are the charge information of eight signals from a PET detector block.To evalute the peformance of this prototype, we adjust the amplitues of the input signals, and equivalently obtained 9 positions scattered over one detector block (including 4 corners, one center, and four position points in the middle of the four edges). The test results are shown in figure 14, which indicate that a position resolution of about 3.3‰ FWHM is achieved.

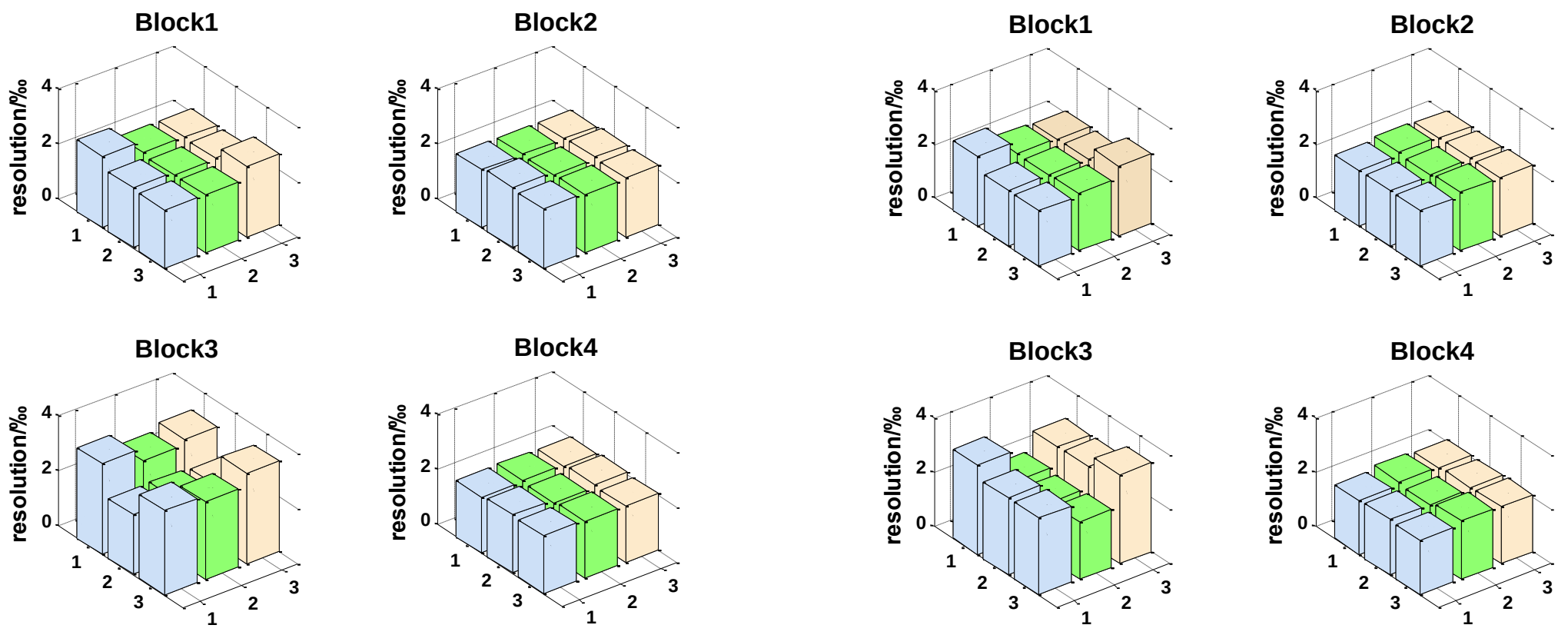

(a) X direction (b) Y direction

Figure 14. Position resolution of 9 typical points for all 4 blocks.

### 3.2 Combination tests with PET detector

We also connect the SPU prototype with the PET detector and conducted tests. Figure 15(a) shows the system under test, and figure 15(b) shows interconnection betwen the FAB and the PET detector.

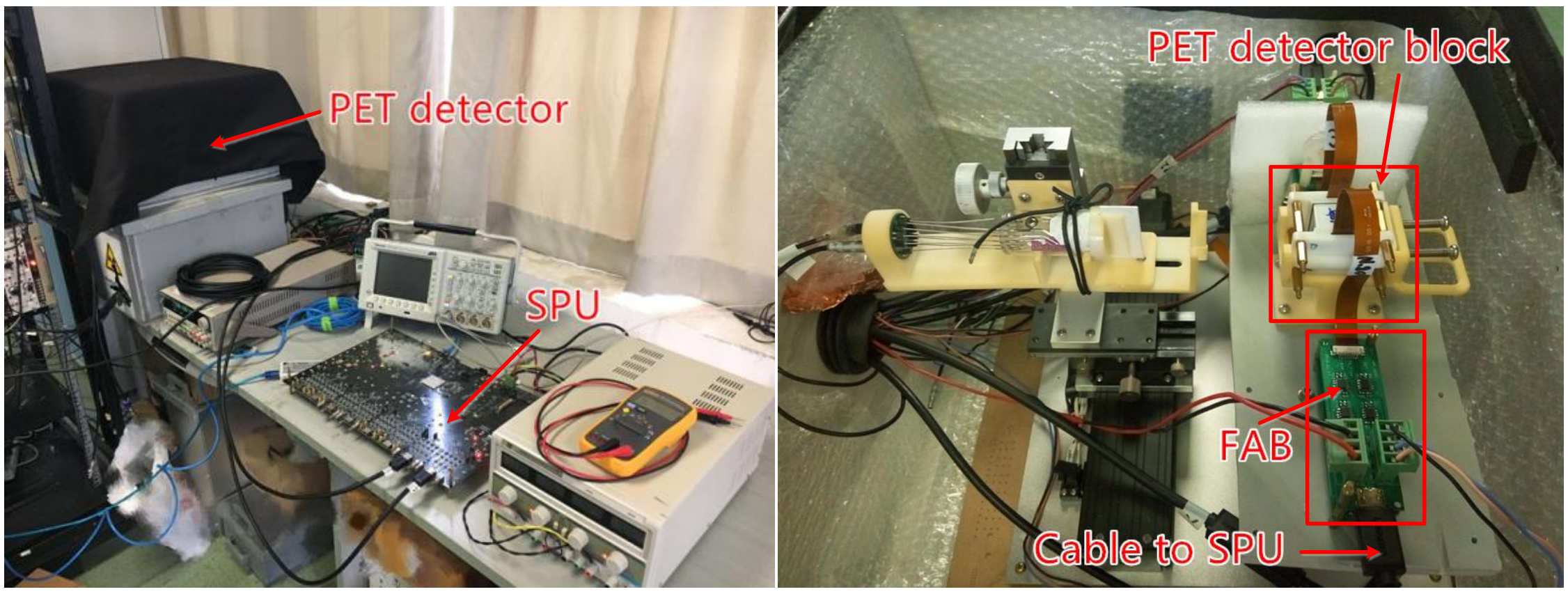

(a) (b)

Figure 15. Pictures of the test system.

As mentioned above, a PET detector block consists of a 23×23 LYSO crystals array. Figure 16 shows one typical flood histogram of one PET detector block, and it indicates that with this SPU prototype, the 23×23 LYSO crystals can be resolved clearly.

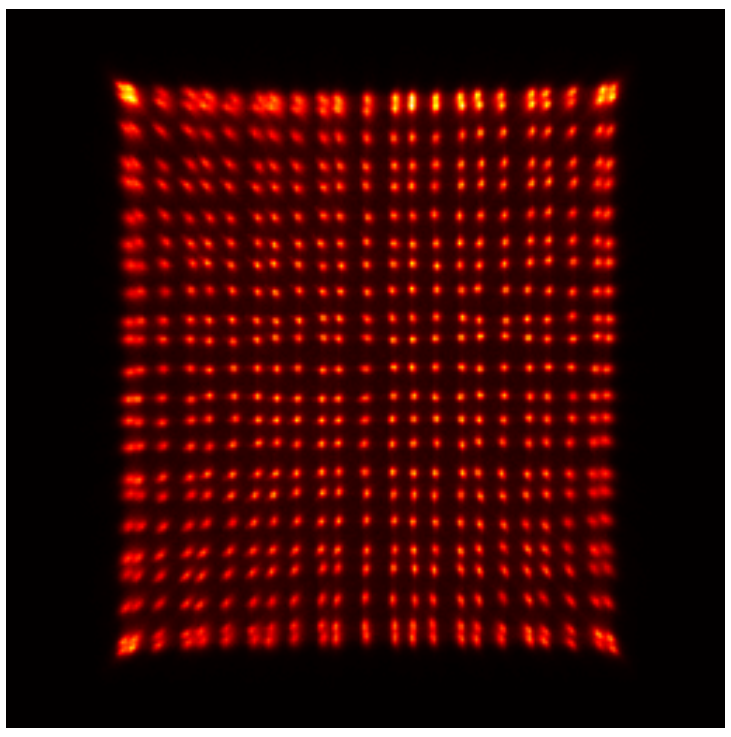

Figure 16. Flood histogram of one PET detector block.

Figure 17 shows the energy spectra obtained with a large amount of events. It indicates that peak energy is located exactly at 511 keV as expected, and the average energy resolution is around 10.2%.

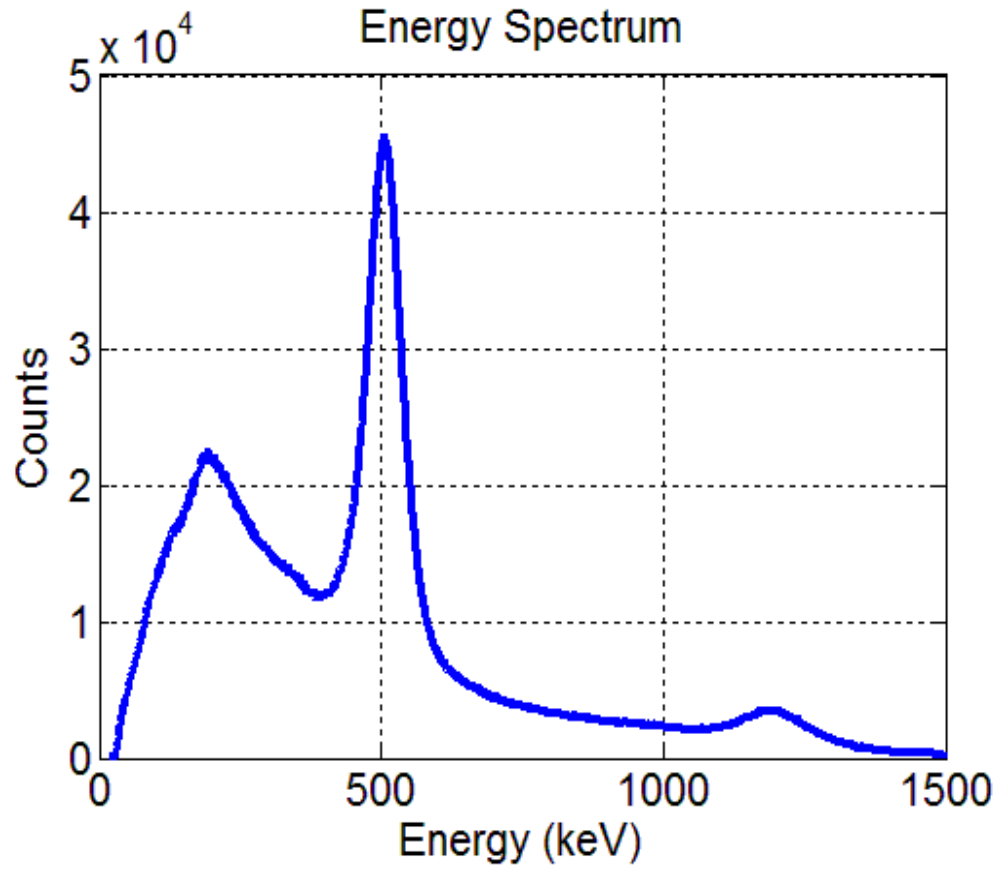

Figure 17. Energy spectra of all craystals.

Figure 18 shows the time measurement results of two PET detector blocks connected to the SPU. It indicates that the time precision is 2.4 ns FWHM, which also meets the application requirement.

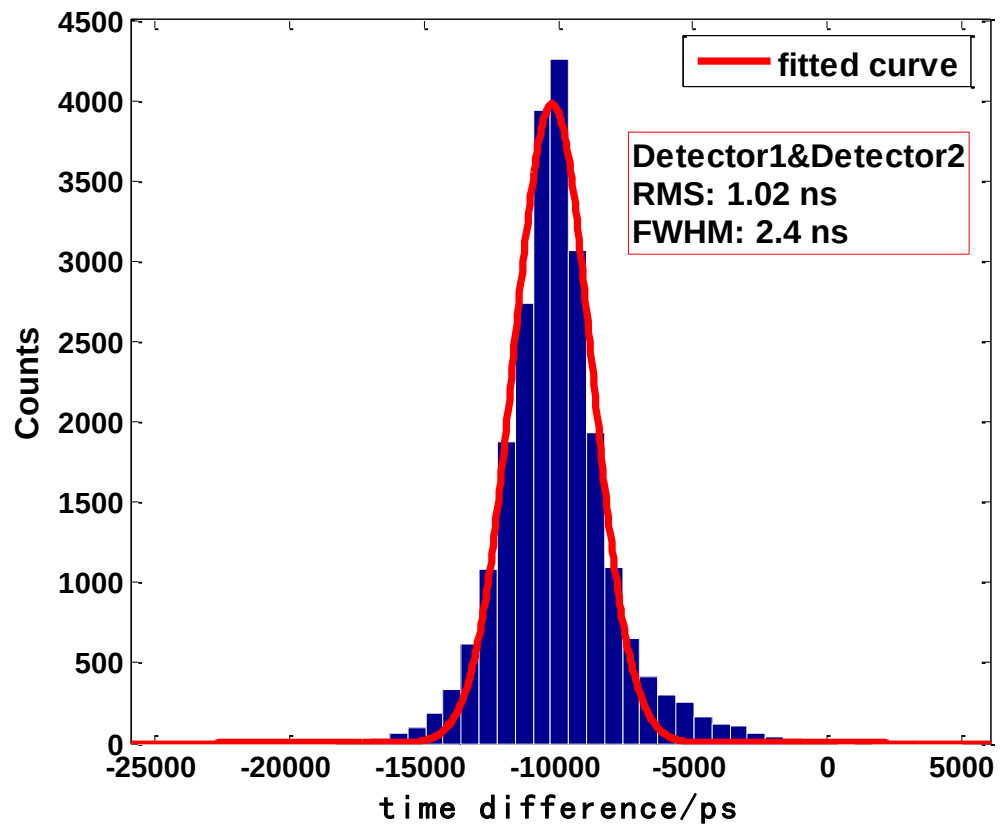

Figure 18. Time measurement histogram of two PET detector blocks.

## 4. Conclusion

A prototype of SPU for a small animal PET system was designed. Through analog signal amplification, shaping, A/D conversion and the real-time area calculation logic in the FPGA, the charge information is obtained, while time measurement is achieved based on analog signal summation, discrimination and Time-to-Digital conversion.

Initial tests have also been conducted. The results indicate that the charge and time precision is better than 3‰ FWHM and 350 ps FWHM respectively, while the position resolution is better than 3.5‰ FWHM. Combination tests of the SPU prototype with the PET detector indicate that the system time precision is better than 2.5 ns, while the flood map and energy spectra concored well with the expected.

## Acknowledgments

We thank Yongfeng Yang, Yibao Wu, Xiaohui Wang, Zhonghua Kuang, and the other collaborators in ShenZhen Institutes of Advanced Technology, CAS for their constant help in our SPU design work.

This work was supported in part by the Knowledge Innovation Program of the Chinese Academy of Sciences under Grant KJCX2-YW-N27 and in part by the CAS Center for Excellence in Particle Physics (CCEPP).

## References

[1] M.E. Phelps, et al. *Application of annihilation coincidence detection to transaxial reconstruction tomography*, 1975 *Journal of Nuclear Medicine Official Publication Society of Nuclear Medicine* **16** 210.

[2] P.J. Slomka, et al. *Recent Advances and Future Progress in PET Instrumentation*, 2016 *Seminars in Nuclear Medicine* **46** 5.

[3] P.M. Bloomfield, et al. *The design and physical characteristics of a small animal positron emission tomograph*, 1995 *Physics in Medicine & Biology* **40** 1105.

[4] S.R. Cherry, et al. *MicroPET: a high resolution PET scanner for imaging small animals*, 1997 *IEEE Transactions on Nuclear Science* **44** 1161.

[5] H. Alva-Sánchez, et al. *A small-animal PET system based on LYSO crystal arrays, PS-PMTs and a PCI DAQ board*, 2010 *IEEE Transactions on Nuclear Science* **57** 85.

[6] J. Wehner, et al. *PET/MRI insert using digital SiPMs: Investigation of MR-compatibility*, 2014 *Nuclear Instruments and Methods in Physics Research Section A: Accelerators, Spectrometers, Detectors and Associated Equipment* **734** 116.

[7] Z. Kuang, et al. *Performance of a high-resolution depth encoding PET detector using barium sulfate reflector*, 2017 *Physics in Medicine & Biology* **62** 5945.

[8] A.L. Lehnert, et al. *Depth of Interaction Calibration and Capabilities in 2×2 Discrete Crystal Arrays and Digital Silicon Photomultipliers*, 2016 *IEEE Transactions on Nuclear Science* **63** 4.

[9] J.D. Thiessen, et al. *MR-compatibility of a high-resolution small animal PET insert operating inside a 7 T MRI*, 2016 *Physics in Medicine & Biology* **61** 7934.

[10] S. Yamamoto, et al. *Development of a Si-PM-based high-resolution PET system for small animals*, 2010 *Physics in Medicine & Biology* **55** 5817.

[11] S.I. Kwon, et al. *Development of small-animal PET prototype using silicon photomultiplier (SiPM): initial results of phantom and animal imaging studies*, 2011 *Journal of Nuclear Medicine* **52** 572.

[12] ADI Corporation, AD8099 data sheet [Online].

http://www.analog.com/media/en/technical-documentation/data-sheets/AD8099.pdf

[13] ADI Corporation, AD8057 data sheet [Online].

http://www.analog.com/media/en/technical-documentation/data-sheets/AD8057_8058.pdf

[14] ADI Corporation, AD8137 data sheet [Online].

http://www.analog.com/media/en/technical-documentation/data-sheets/AD8137.pdf

[15] ADI Corporation, AD9222 data sheet [Online].

http://www.analog.com/media/en/technical-documentation/data-sheets/AD9222.pdf

[16] TI Corporation, LMH7322 data sheet [Online].

http://www.ti.com/lit/ds/symlink/lmh7322.pdf

[17] C. Ye, et al. *A field-programmable-gate-array based time digitizer for the time-of-flight mass spectrometry*, 2014 *Review of Scientific Instruments* **85** 045115.

[18] L. Zhao, et al. *A 16-channel high-resolution time and charge measurement module for the external target experiment in the CSR of HIRFL*, 2014 *Nuclear Science and Techniques* **25** 010401.